%% file: optdmc_journal.tex
\documentclass[draftclsnofoot, onecolumn, letterpaper, romanappendices]{IEEEtran}
\IEEEoverridecommandlockouts
\usepackage[pdftex]{graphicx}
\usepackage{epstopdf}
\usepackage{psfrag,graphicx}
\usepackage{verbatim}
\usepackage{amsfonts}
\usepackage{caption}
\usepackage{amssymb}
\usepackage{amsmath}
\usepackage{amsthm}
\input defs.tex

\begin{document}

\newcommand*\samethanks[1][\value{footnote}]{\footnotemark[#1]}
\newtheorem{Theorem}{\textbf{Theorem}}
\theoremstyle{lemma}
\newtheorem{lemma}{\textbf{Lemma}}
\newtheorem{corollary}{\textbf{Corollary}}
\theoremstyle{definition}
\newtheorem*{Proof}{\textbf {Proof}}

\newenvironment{definition}[1][Definition]{\begin{trivlist}
\item[\hskip \labelsep {\bfseries #1}]}{\end{trivlist}}
\newenvironment{example}[1][Example]{\begin{trivlist}
\item[\hskip \labelsep {\bfseries #1}]}{\end{trivlist}}
\newenvironment{remark}[1][Remark]{\begin{trivlist}
\item[\hskip \labelsep {\bfseries #1}]}{\end{trivlist}}

\title{No input symbol should occur more frequently than $1-\frac{1}{e}$}

\author{\authorblockN{Gowtham Kumar*,   \thanks{*Stanford University. Email: amanolak@stanford.edu, gowthamr@stanford.edu. }}  \and \authorblockN{Alexandros Manolakos* } }

\maketitle 

\begin{abstract}
Consider any discrete memoryless channel (DMC) with arbitrarily but finite input and output alphabets $\mathcal{X}$, $\mathcal{Y}$ respectively. Then, for any capacity achieving input distribution all symbols occur less frequently than $1-\frac{1}{e}$. That is,
\[
\max\limits_{x \in \mathcal{X}} P^*(x)  < 1-\frac{1}{e}
\]
\noindent where $P^*(x)$ is a capacity achieving input distribution. Also, we provide sufficient conditions for which a discrete distribution can be a capacity achieving input distribution for some DMC channel. Lastly, we show that there is no similar restriction on the capacity achieving output distribution.

\end{abstract}


\section{Introduction}
For an arbitrary discrete probability distribution, under what circumstances can we find a
discrete memoryless channel (DMC) for which the given distribution achieves the
capacity of that channel?  Is it always possible to find a channel for any
arbitrary distribution?  As surprising as it might seem, the answer is negative. That is, there exist probability distributions that can never be capacity achieving distributions for any discrete memoryless channel.  More precisely, the main result of this work is that a source distribution that transmits a symbol with probability greater than or equal to $1-\frac{1}{e}$ can never be a capacity achieving distribution. 

The result stated above, leads to the following natural question. Is there a similar restriction on the capacity achieving output distribution of a discrete memoryless channel? Using a dual characterization of the channel capacity we are going to argue that all probability distributions can be capacity achieving output distributions for some channel.

Last but not least, we asked whether there exist simple sufficient conditions on whether an arbitrary probability distribution is a capacity achieving distribution for some channel. Consider an input probability distribution $P(x)$. If there is a subset of symbols whose sum of probabilities lies in the  ($\frac{1}{e}$, $1-\frac{1}{e}$) interval, then there exists a channel for which the distribution $P(x)$ is capacity achieving.

The paper is organized as follows. In section \ref{Preliminaries} we  review some important definitions and we introduce our notation. In section \ref{binaryIO} we present the starting point of this work and in section \ref{multipleO} we extend the latter result to the case of discrete memoryless channels with two inputs and multiple outputs. Then, in section \ref{multipleIO} we present the most general result for an arbitrary discrete memoryless channel with multiple inputs and multiple outputs and we state a dual information geometric result. Lastly, in section \ref{converse} we present sufficient conditions for an input distribution to be capacity achieving for a DMC.
\section{Preliminaries}
\label{Preliminaries}

 We require the following definitions \cite{cover1991}.

\begin{definition}
A discrete channel, denoted by $(\mathcal{X},P(y|x),\mathcal{Y})$, consists of two finite sets $\mathcal{X}$ and $\mathcal{Y}$ and a collection of probability mass functions $P(y|x)$, one for each $x \in \mathcal{X}$, such that for every $x$ and $y$, $P(y|x) \geq 0$, and for every $x$, $\sum\limits_{y \in \mathcal{Y}} P(y|x) =1$. 
\\A channel is memoryless if 
\begin{align}
P(y^n|x^n) & = \prod\limits_{i=1}^n P(y_i|x_i)
\end{align}
\noindent for all $n \geq 1$.
\end{definition}
\begin{definition}
We define the information channel capacity of a discrete memoryless channel $(\mathcal{X},P(y|x),\mathcal{Y})$  as
\begin{align}
C & = \max\limits_{P(x)} I(X ; Y)
\end{align}
\noindent where $I(X;Y)$ denotes the mutual information and the maximum is taken over all possible input distributions $P(x)$.
 \end{definition}
 Denote 
\begin{align}
P^*(x)= \arg\max\limits_{P(x)} I(X ;Y) \\
Q^*(y)=\sum\limits_{i=1}^{|\mathcal{X}|}P(y|x_i)P^*(x_i)
\end{align}
the capacity achieving input distribution and the capacity achieving output distribution respectively.

Note that there may exist more than one capacity achieving input distribution for a given channel, but they all induce the same capacity achieving output distribution. Also, we do not consider trivial channels with capacity zero. 

The capacity of a discrete memoryless channel can be calculated using the following dual representation due to Gallager and Ryabko \cite{cover1991}. Consider a DMC $(\mathcal{X},P(y|x),\mathcal{Y})$. The capacity is given by
\begin{align}
C &= \min\limits_{Q} \max\limits_{x \in \mathcal{X}} D( P(\cdot | x) || Q(y) ).
\end{align}

\begin{figure}
\begin{center}
\includegraphics[height=1.5in, width=2in]{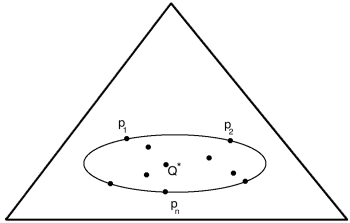}
 \caption{ Minimum radius Information ball. \label{fig5}}
 \end{center}
\end{figure}

\noindent This means that the capacity of a channel can be interpreted as the radius of the smallest information ball containing the rows of $P(\cdot | x),~x \in \mathcal{X}$. Then the minimizing $Q$ is the center of this ball and $Q$ is the capacity achieving output distribution.

\section{Binary Input, Binary Output}
\label{binaryIO}
\begin{figure}[h!]
\begin{center}
\includegraphics[width=0.25\columnwidth]{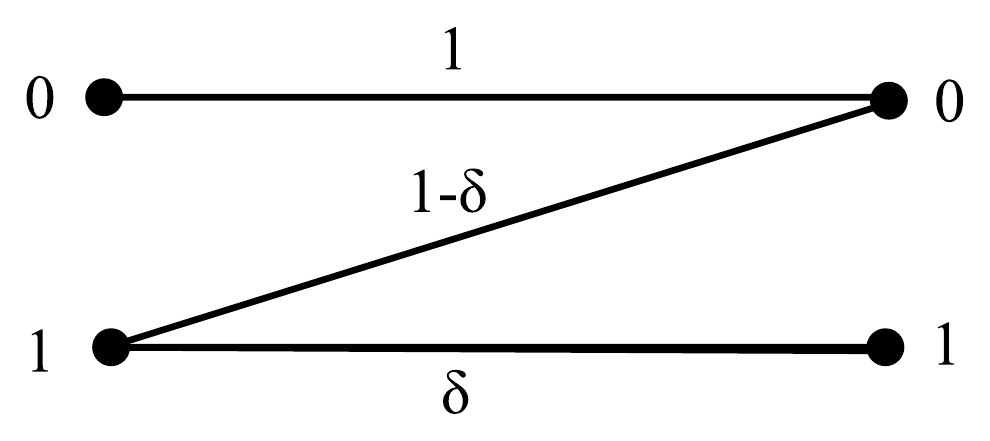}
 \caption{Z-Channel with capacity approaching zero.\label{fig1}}
\end{center}
\end{figure}

For the binary-input binary-output channel it is possible to obtain a simple 
analytical solution for the capacity achieving distribution in terms of the
transition probabilities \cite{Silverman}. Using these  formulas, it turns out that for a binary channel, neither symbol
should be transmitted with probability greater than or equal to $1-\frac{1}{e}$ in order to achieve capacity. Note that there is no DMC for which the distribution $P_{0}(x)$

\begin{eqnarray}
P_{0}(x) = \begin{cases} \frac{1}{e} &\mbox{if } x=0\\
1-\frac{1}{e} &\mbox{if } x=1
\end{cases}
\end{eqnarray}

\noindent is capacity achieving. However, there exists the Z-channel of Fig. \ref{fig1} with capacity achieving distribution

\begin{eqnarray}
P(x) = \begin{cases} \frac{1}{e}+\epsilon(\delta) &\mbox{if } x=0\\
1-\frac{1}{e}-\epsilon(\delta) &\mbox{if } x=1
\end{cases}
\end{eqnarray}

\noindent that has capacity $C=C(\delta)$, where $C(\delta) \rightarrow 0$ and $\epsilon(\delta) \rightarrow 0$ as $\delta \rightarrow 0$. Therefore, $P_{0}(x)$ is never an optimal input distribution for a non-trivial discrete memoryless channel.

\noindent Next, we generalize this result for the case of a DMC with $|\mathcal{Y}| \geq 2$.

\section{Binary Input, Multiple Output}
\label{multipleO}
In this section, we will prove that increasing the output alphabet size for a binary input does not change our restriction on the input symbol
probabilities needing to belong in the interval $(\frac{1}{e}, 1-\frac{1}{e})$.

\begin{lemma}
\label{lemma4}
The following function
\begin{align}
f(\alpha;p_1,p_2)&=
p_1\log \frac{p_1}{\alpha p_1+\bar{\alpha} p_2}-p_2\log \frac{p_2}{{\alpha}p_1+\bar{\alpha} p_2} -\left(p_1-p_2\right)
\label{eqtr1}
\end{align}
\noindent where $\bar{\alpha}=1-\alpha$ and $p_1,p_2,\alpha \in [0,1]$ satisfies
\begin{align}
f(\frac{1}{e};p_1,p_2) \geq 0
\end{align}
\noindent with equality iff $p_1=p_2$ or $p_2=0$.
\end{lemma}

\begin{Proof}
See Appendix.
\end{Proof}

\begin{Theorem}
\label{input1}
Let $(\mathcal{X},P(y|x),\mathcal{Y})$ be  a two input discrete memoryless channel with transition matrix  $\left [ \begin{array}{c} P_1(y) \\  P_2(y) \end{array} \right ] $, where $P_i(j)=P(y=j|x=i)$ corresponds to the conditional probability of receiving symbol $j$ given that symbol $i$ is
transmitted. We assume that $|\mathcal{Y}| \geq 2$. Let $\alpha^*:=\arg\max\limits_{0 \leq \alpha \leq 1} I(X;Y)$, where $\alpha=\text{Pr}\{X=1\}$. \\ Then
\begin{align}
\frac{1}{e} < \alpha^* < 1- \frac{1}{e}.
\end{align}
\end{Theorem}

\begin{figure}[h!]
\begin{center}
    \includegraphics[height=0.3\linewidth]{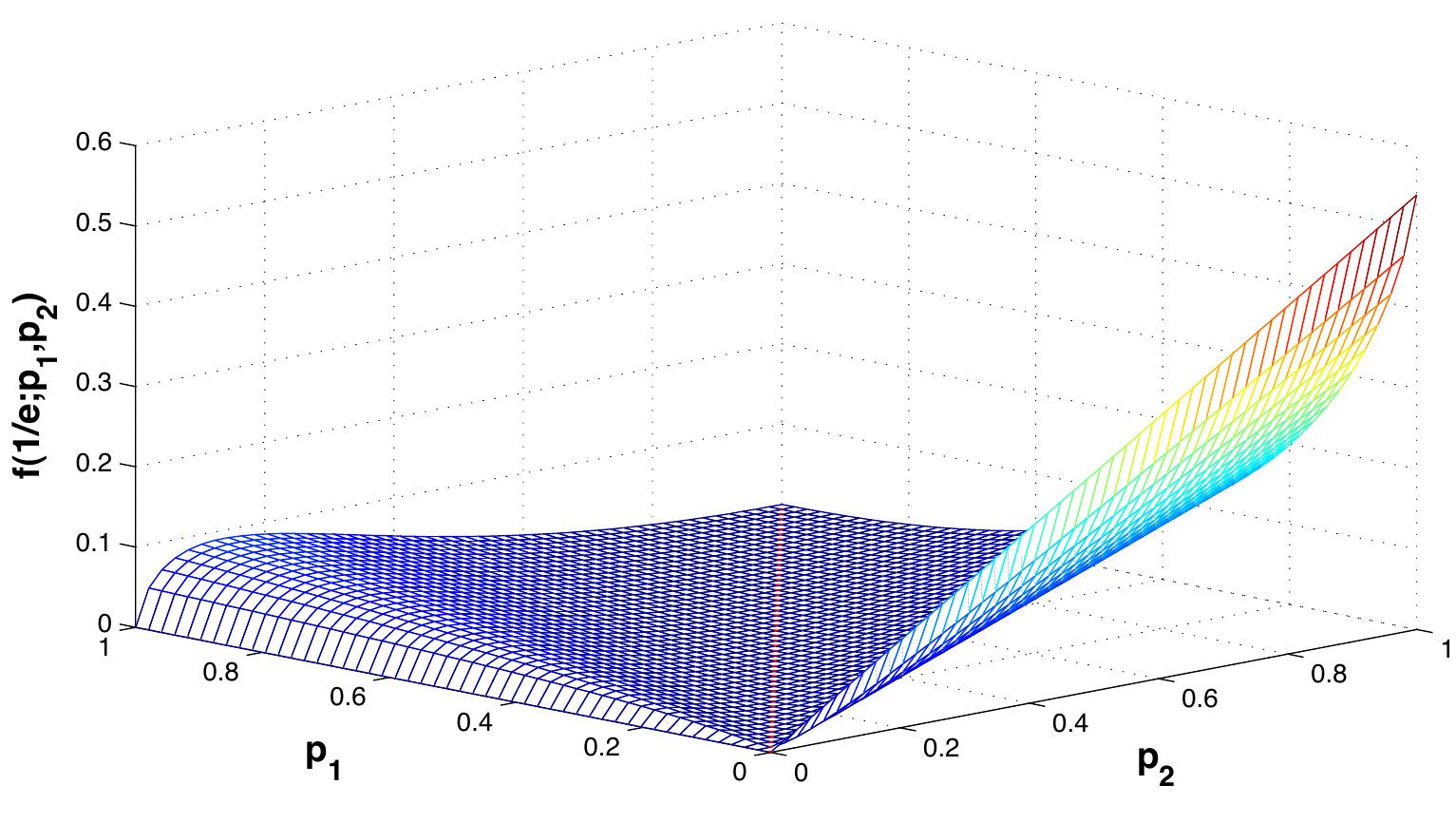}
\caption{ $f(\frac{1}{e};p_1,p_2) \geq 0$  for $p_1,p_2 \in [0,1]$.\label{fig45}}
\end{center}
\end{figure}

\begin{Proof}The capacity $C$ of the DMC is
given by:
\begin{align}
C & = \max\limits_{P(x)} I(X;Y) \\
& = \max\limits_{\alpha}  I(X;Y)
\end{align}
\noindent where the maximization is taken over all input distributions $P(x) =
\Pr\{X=x\},x \in \{0,1\}$, $\alpha=\text{Pr}\{X=1\}$ and $\bar{\alpha}=1-\alpha$.\\
\noindent Note that since $I(X;Y)$ is a concave function of $\alpha$
\begin{align}
\frac{ dI(X;Y)}{d \alpha} = D(P_1 || Q_{\alpha}) - D(P_2 || Q_{\alpha})
\end{align}
\noindent is a non-increasing function of $\alpha$. It suffices then to prove that 
\begin{align}
\frac{ dI(X;Y)}{d \alpha} \Big |_{\alpha=\frac{1}{e}} >0 \text{ and } \frac{ dI(X;Y)}{d \alpha} \Big |_{\alpha=1-\frac{1}{e}} <0
\end{align}
\noindent because in that case by the intermediate value theorem the solution $\alpha^*$ of $\frac{ dI(X;Y)}{d \alpha}  = 0$ satisfies 
\begin{align}
\nonumber
\alpha^* \in (\frac{1}{e},1-\frac{1}{e}).
\end{align}

\noindent To this end, we have that:
\begin{align}
I(X;Y)  & =  D(P(x)P(y)||P(x,y)) \\
& = \sum\limits_{y=1}^{|\mathcal{Y}|} \sum\limits_{x \in \mathcal{X}} P(x) P(y|x) \log(\frac{P(y|x)}{P(y)}) \\
& =  \sum\limits_{y=1}^{|\mathcal{Y}|} \Big [  \alpha P_1(y) \log(\frac{P_1(y)}{\alpha P_1(y)+\bar{\alpha}P_2(y)}) +\bar{\alpha}P_2(y) \log(\frac{P_2(y)}{\alpha P_1(y)+\bar{\alpha} P_2(y)}) \Big ] 
\end{align}
\noindent  Note that 
\begin{align}
\frac{ dI(X;Y)}{d \alpha} = D(P_1 || Q_{\alpha}) - D(P_2 || Q_{\alpha})
\end{align}
\noindent where $Q_{\alpha}(y) = \alpha P_1(y) + \bar{\alpha} P_2(y)$ and that 
\begin{align}
D(P_1 || Q_{\alpha}) - D(P_2 || Q_{\alpha}) = \sum\limits_{y=1}^{|\mathcal{Y}|} f(\alpha ;  P_1(y), P_2(y))
\end{align}
\noindent where $f$ is defined in Eq. \ref{eqtr1}.
\noindent Notice that by Lemma \ref{lemma4}
\begin{align}
\sum_{y=1}^{|\mathcal{Y}|} f(\frac{1}{e}; P_1(y), P_2 (y))& \geq 0
\end{align}
\noindent with equality iff $P_1(y)=P_2(y),~\forall~y \in \mathcal{Y}$. When $P_1(y)=P_2(y),~\forall~y \in \mathcal{Y}$, $C=0$ is a trivial case that we ignore. Therefore,
\begin{align}
\nonumber
\sum_{y=1}^{|\mathcal{Y}|} f(\frac{1}{e}; P_1(y), P_2 (y)) > 0 \\ \Rightarrow
D(P_1||Q_{\frac{1}{e}})-D(P_2||Q_{\frac{1}{e}}) > 0
\label{eqrt3}
\end{align}

\noindent By interchanging $P_1$ and $P_2$ above we get that
\begin{align}
D(P_1||Q_{1-{1\over e}})-D(P_2||Q_{1-{1\over e}}) < 0
\label{eqrt4}
\end{align}
\noindent From Eq. (\ref{eqrt3}) and (\ref{eqrt4}) it follows that
\begin{align}
\frac{ dI(X;Y)}{d \alpha} \Big |_{\alpha=\frac{1}{e}} >0 \text{ and } \frac{ dI(X;Y)}{d \alpha} \Big |_{\alpha=1-\frac{1}{e}} <0.
\end{align}
\noindent Therefore by the intermediate value theorem the solution $\alpha^*$ of $\frac{ dI(X;Y)}{d \alpha}  = 0$ satisfies
\begin{align*}
\nonumber
\alpha^* \in (\frac{1}{e},1-\frac{1}{e}).
\end{align*}
This completes the proof. \qed
 \end{Proof}
 
\begin{corollary}
\label{input2}
Let $P_1(y)$ and $P_2(y)$ be any two probability distributions on $\mathcal{Y}$.  Let $\alpha \in[0,1]$ be chosen such that $D(P_1||Q_\alpha)=D(P_2||Q_\alpha)$ where $Q_\alpha(y)=\alpha P_1(y)+(1-\alpha)P_2(y)$. Then $\alpha \in(\frac{1}{e},1-\frac{1}{e})$, where $e$ is the base of the natural logarithm.
\end{corollary}

\begin{figure}
\begin{center}
    \includegraphics[height=0.25\linewidth]{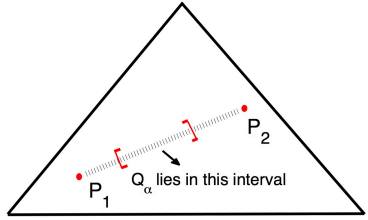}
\caption{Geometric interpretation of Corollary \ref{input1}.\label{fig4}}
\end{center}
\end{figure}

\noindent In Fig. \ref{fig4} we show a geometric representation of Corollary \ref{input1}. For any two distributions $P_1$ and $P_2$, the distribution $Q_{\alpha}$ that satisfies $D(P_1||Q_\alpha)=D(P_2||Q_\alpha)$ lies in the interval as shown in the figure.

\begin{corollary}
\label{lemma1}
Let $P_1(y)$ and $P_2(y)$ be any two probability distributions on $\mathcal{Y}$ ($P_1 \neq P_2$). Let  $ Q_{\frac{1}{e}}(y)=\frac{1}{e} P_1(y)+ (1-\frac{1}{e})P_2(y)$. Then,
\begin{equation}
D(P_1||Q_{1\over e})-D(P_2||Q_{1\over e}) > 0. \\
\end{equation}
\end{corollary}

\noindent Corollary \ref{lemma1} has an interesting interpretation in an information geometric sense. Specifically, consider any two distributions $P_1$ and $P_2$. Then the distribution $Q_\alpha$ lies on the line  segment that connects the distributions $P_1$ and $P_2$. The farthest that $Q_\alpha$ can be from $P_2$ and still always be closer to $P_2$ than to $P_1$ is for $\alpha=\frac{1}{e}$. 

\begin{corollary}
\label{lemma3}
 Consider a cost constraint $l(x)$ on the input symbols such that $l(0)=0$ and $l(1)=1$. Let $\alpha=Pr(x=1)$. Define the following problem:
\BEQ
\begin{array}{ccc}
C_{\rho} = \max \limits_{\alpha \in [0,1]} & I(X;Y)  \\
\mbox{subject to} &   \mathbf{E}l(x) \leq  \rho \\
\label{eq34}
\end{array}
\EEQ
\noindent If $\rho \leq \frac{1}{e}$ then $\alpha^*=\rho$.
\end{corollary}
\noindent As an example of the application of Corollary \ref{lemma3} consider a binary DMC where the fraction of $1$'s is constrained to be at most say $20$\%. According to Theorem 1 the fraction of $1$'s is at least $\frac{1}{e}$ (around 36\%). Then from Corollary 3 we conclude that the capacity achieving fraction of $1$'s is exactly $20$\%.

\section{Maximum Symbol Probability for a Multiple Input, Multiple Output DMC}
\label{multipleIO}

  We now expand the above result to show that in fact, for any capacity achieving input distribution of a discrete memoryless channel, no input symbol  can ever have a probability greater than or equal to $1 - \frac{1}{e}$.

\begin{Theorem}
\label{mainth}
Let any discrete memoryless channel (DMC) $(\mathcal{X},P(y|x),\mathcal{Y})$   with input alphabet $\mathcal{X}:=\{0,1,2, \dots m-1\}$ and output alphabet $\mathcal{Y}:=\{0,1,2, \dots n-1\}$. Let $P(x)$ be the input distribution over the alphabet $\mathcal{X}$. \\ Define $P^*(x):=\arg\max\limits_{P(x)} I(X;Y)$. Then
\begin{align}
\max\limits_{ 0 \leq x \leq m-1 } P^*(x) < 1-\frac{1}{e}
\end{align}
\end{Theorem}

\begin{Proof}
To show that the $1 - \frac{1}{e}$ result extends beyond the binary input case,  without loss of generality  we  need to prove that $\text{Pr}\{X=0\} < 1-\frac{1}{e}$. Define a function $f(x)$ on our input

\begin{eqnarray}
f(x) = \mathbf{1}\{x \neq 0\} = \begin{cases} 0 &\mbox{if } x = 0\\
1 &\mbox{if } x \neq 0
\end{cases}
\end{eqnarray}

\noindent Denote $\Pr\{f(x)=0\}=\alpha$. Also, denote $\bar{\alpha}=1-\alpha$.  \\The capacity $C$ of the channel is given by
\begin{eqnarray}
\nonumber
C &=& \max\limits_{P(x)} I(X;Y) \\
\nonumber
&\stackrel{(a)}{=}& \max\limits_{P(x,f(x))} I(X,f(X);Y) \\
\nonumber
  &=& \max\limits_{P(x,f(x))} [I(f(X);Y) + I(X;Y|f(X))] \\
\nonumber
   &\stackrel{(b)}{=}& \max\limits_{P(x,f(x))} [I(f(X);Y) + \bar{\alpha} I(X;Y|f(X)=1)] \\
   \nonumber
   &\stackrel{(c)}{=}& \max\limits_{P(x|f(x))} \max\limits_{P(f(x))}  [ I(f(X);Y) + \bar{\alpha} I(X;Y|f(X)=1)  ] \\
   & = &  \max\limits_{P(x|f(x))} \max\limits_{\alpha}  [ I(f(X);Y) + \bar{\alpha} I(X;Y|f(X)=1)  ] 
\end{eqnarray}
\noindent where
\begin{itemize}
\item $(a)$ follows from the fact that $f(X)$ is a deterministic function of $X$.
\item $(b)$ follows from the fact that when $f(X) = 0$, $H(X|f(X)=0) = 0$
and thus $I(X;Y|f(X)=0)=0$.
\item $(c)$ follows because the choice of $P(f(x))$
can be made independent of $P(x|f(x))$, so we can split the maximization.
\end{itemize}

\noindent Let $\alpha_1= \arg\max\limits_{\alpha} I(f(X);Y)$. Fix  $P(x|f(x))$. Then, from Theorem \ref{input1}
\begin{align}
\alpha_1 < 1-\frac{1}{e}
\label{eq28}
\end{align}
\noindent and we also know that 
\begin{align}
\label{eq29}
\left . I(f(X);Y) \right |_{\alpha=\alpha_1} & \geq \left . I(f(X);Y) \right |_{\alpha>\alpha_1} \\
\label{eq30}
\left . \bar{\alpha} I(X;Y|f(X)=1) \right |_{\alpha=\alpha_1} & \geq \left . \bar{\alpha} I(X;Y|f(X)=1) \right |_{\alpha>\alpha_1}
\end{align}
\noindent The first inequality follows from the definition of $\alpha_1$ and the second inequality follows easily because \\ $\bar{\alpha} I(X;Y|f(X)=1)$  is a linear function of $\alpha$.\\
\noindent Adding inequalities (\ref{eq29}) and (\ref{eq30}) we get that:
\begin{align}
\label{eq31}
[ I(f(X);Y) &+ \bar{\alpha} I(X;Y|f(X)=1) ]  |_{\alpha=\alpha_1} \geq [ I(f(X);Y) + \bar{\alpha} I(X;Y|f(X)=1) ]  |_{\alpha>\alpha_1}
\end{align}

\noindent From inequality (\ref{eq31}) we conclude that
\begin{align}
\arg\max\limits_{\alpha} [I(f(X);Y) + \bar{\alpha} I(X;Y|f(X)=1)] \leq \alpha_1  \\
\stackrel{(\ref{eq28})}{\Rightarrow} \arg\max\limits_{\alpha} [I(f(X);Y) + \bar{\alpha} I(X;Y|f(X)=1)] < 1-\frac{1}{e}
\end{align}
\noindent Note that the latter holds for any $P(x|f(x))$. \\ Thus, 
\begin{align}
\alpha = \text{Pr}\{X=0\} < 1-\frac{1}{e}.
\end{align}
This completes the proof. \qed
\end{Proof}

\begin{figure}[h!]
\begin{center}
\centering
\includegraphics[width=0.35\columnwidth]{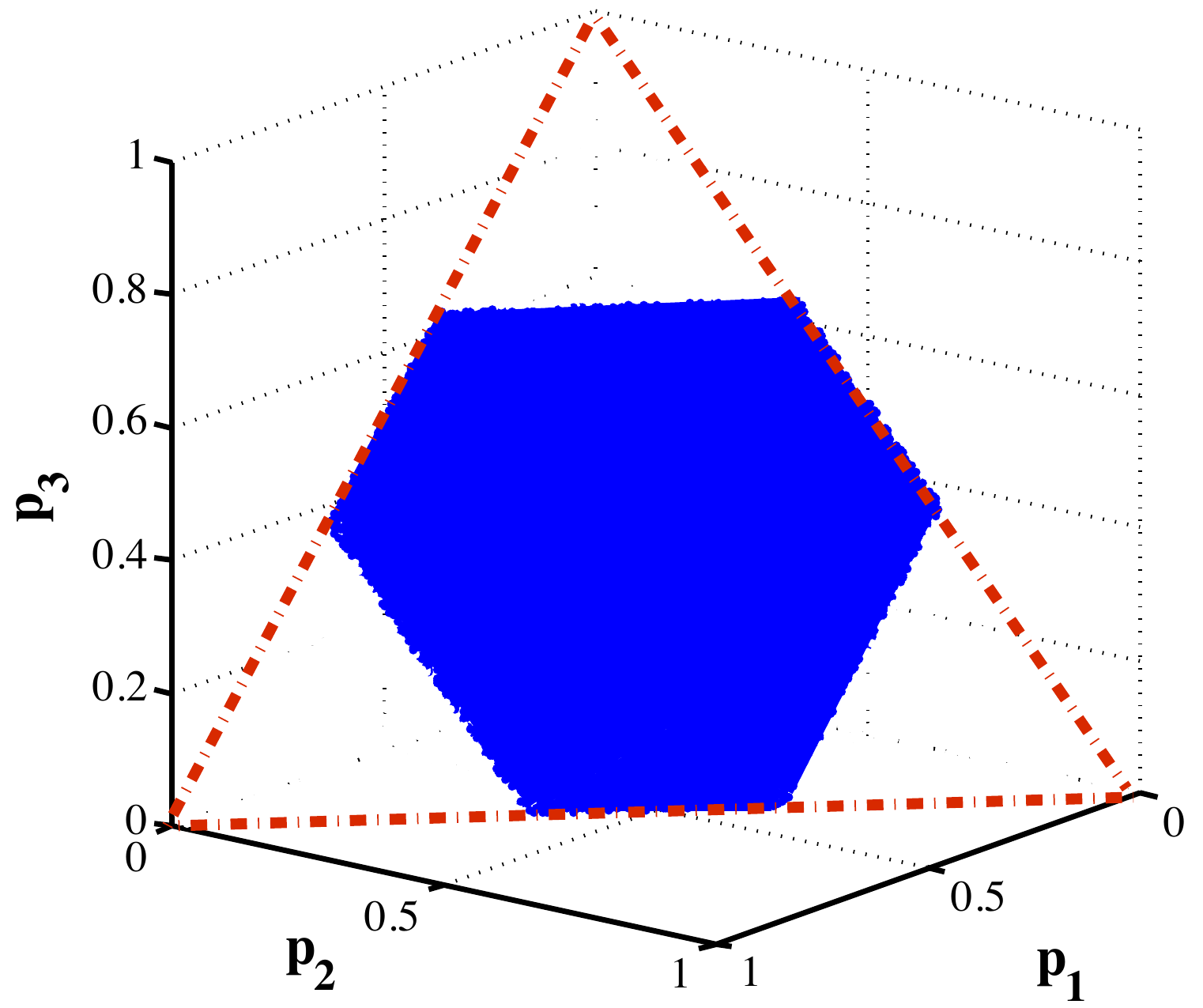}
\caption{Space of optimal input distributions for a 3-3 DMC.\label{fig2}}
\end{center}
\end{figure}

\noindent In Fig. \ref{fig2} we show the region of optimal input  distributions  $P^*(x)$ (dark region) on the $3$-simplex that satisfy the constraint  $\max\limits_{ 0 \leq x \leq m-1 } P(x) < 1-\frac{1}{e}$.

\noindent The above result induces the following constraints for the capacity achieving output distribution.
\begin{corollary}
\label{cor2}
Let $P_1(y),P_2(y),P_3(y),\dots, P_m(y)$ be any $m$ distributions on $\mathcal{Y}$. Let $Q^*(y)$ be the center of the information ball of minimum radius that contains the distributions $P_1,P_2,\dots,P_m$ on the $\mathcal{Y}$-simplex.  \\That is, let $Q^*(y)$ be the solution of the optimization problem
\BEQ
\begin{array}{ccc}
\medskip
\min \limits_{Q(y)}  & C\\
\medskip
\mbox{subject to} &  D(P_i || Q) \leq C,~ 1\leq i \leq m \\
\medskip 
& Q(y)= \sum\limits_{i=1}^m \alpha_i P_i(y) \\
\medskip

& \sum\limits_{i=1}^m \alpha_i = 1 \\
\end{array}
\EEQ
\noindent Then $\alpha_i < 1-\frac{1}{e},~ 1\leq i \leq m $.\\
\end{corollary}
\begin{Proof}
By the min-max Duality Theorem for capacity (\cite{cover1991}), $C$ is the capacity of the DMC whose rows are the distributions $P_i$ and $\alpha_i$ is the probability of the input symbol $i$. Applying Theorem \ref{mainth} we get that:
\begin{align*}
\alpha_i < 1-\frac{1}{e}. \qed
\end{align*}
\end{Proof}
\noindent Fig. \ref{fig5} shows a sketch of the region in which the capacity achieving output distribution $Q^*(y)$ lies.  $P_1(y),P_2(y)$ and $P_3(y)$ are the rows of the channel matrix.

\begin{figure}[!h]
\begin{center}
    \includegraphics[height=2in,width=2.8in]{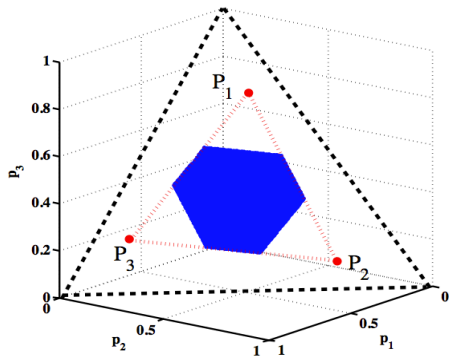}
\caption{ Space of optimal output distributions for a DMC with rows  $P_1(y),P_2(y)$ and $P_3(y)$. \label{fig5}}
\end{center}
\end{figure}

\noindent {\bf Remark.}  The capacity achieving output distribution $Q^*(y)$ can lie anywhere on the $\mathcal{Y}$-simplex. However, given the channel matrix, $Q^*(y)$ lies on the region described by Corollary \ref{cor2} and shown in Fig. \ref{fig5}.

\section{Sufficient Conditions for Optimality}
\label{converse}
So far we have shown that for any discrete memoryless channel with finite input alphabet
$\mathcal{X}$ and finite output alphabet $\mathcal{Y}$, the capacity achieving distribution $P^*(x)$ must satisfy $P^*(x) <
1-\frac{1}{e}$ $\forall x \in \mathcal{X}$.  In this section we show that if there exists a subset of $P(x)$ that sums
in $(\frac{1}{e}, 1-\frac{1}{e})$, then there exists a channel for which this distribution is a capacity achieving input distribution.

\begin{Theorem}
Let $P(x)$ be a discrete input probability distribution over a discrete memoryless channel $(\mathcal{X},P(y|x),\mathcal{Y})$ with  $\mathcal{X}:=\{0,1,2 \dots, m-1\}$. If $\exists~S \subset \mathcal{X} : \sum\limits_{x \in S} P(x) \in (\frac{1}{e}, 1-\frac{1}{e})$ then there exists a channel for which $P(x)$ is a capacity achieving input distribution.
\end{Theorem}

\begin{Proof} We are going to construct a channel for which the distribution $P(x)$ is a capacity achieving input distribution. Let
\begin{align}
 p=\sum\limits_{x \in S} P(x)
 \end{align}
 Then, $p \in (\frac{1}{e}, 1-\frac{1}{e})$. Also, denote $q=1-p$. 
\\ We now construct a $2$-Input, $2$-Output channel with transition matrix
\begin{align}
P_1(y|x)=\left [ \begin{array}{c}  P(y|x=0)\\ P(y|x=1) \end{array}  \right ]
\end{align} such that the distribution

\begin{eqnarray}
Q(x) = \begin{cases} p &\mbox{if } x= 0\\
q &\mbox{if } x=1
\end{cases}
\end{eqnarray}
\noindent is capacity achieving. The latter always exist because for a 2-Input 2-Output DMC, the formulas for the capacity achieving input distribution (\cite{Silverman}) are continuous
over the transition probabilities and $p \in (\frac{1}{e}, 1-\frac{1}{e})$. The latter allows us to conclude that any input distribution which satisfies the $1-\frac{1}{e}$ constraint has a channel for which that distribution is capacity achieving.

\noindent Starting from the transition matrix $P_1(y|x)$, construct a channel in which the second line is cloned $m-2$ times:
\begin{align}
P_2(y|x)=\left [
\begin{array}{c}
P(y|x=0) \\
 P(y|x=1) \\
 P(y|x=1) \\
\vdots \\
 P(y|x=1)
\end{array} \right ]
\end{align}
\noindent Obviously, for this channel any distribution that has
\begin{align}
\label{ert1}
 P(x=0)=p \\ 
 \sum\limits_{i=1}^{m-1} P(x=i)=q
 \label{ert2}
 \end{align} 
 is a capacity achieving input distribution since all the lines of the matrix except the first are exactly the same. Note that the probability distribution $P(x)$ satisfies equations (\ref{ert1}) and (\ref{ert2}) and therefore it is a capacity achieving input distribution for the channel with transition matrix $P_2(y|x)$. This completes the proof. \qed
\end{Proof}


\section{Acknowledgment}
The authors were partially supported by the Center for Science of Information (CSoI), an NSF Science and Technology Center, under grant agreement CCF-0939370 and by Air Force Office of Scientific Research (AFOSR) through grant FA9550-10-1-0124. The authors wish to thank their advisor Thomas Cover for insightful discussions and for his continual encouragement. Also, they would like to  acknowledge very helpful conversations with Kartik Venkat, Bobbie Chern and  Yeow-Khiang Chia.

\section{Appendix}

\noindent {\bf Proof of Lemma \ref{lemma4}.}

\noindent We must prove that $f(\frac{1}{e};p_1,p_2) \geq 0$. To this end, we compute the first and second derivatives of $f(\frac{1}{e}; p_1,p_2)$ over $p_1$:

\begin{align}
\frac{\partial f(\frac{1}{e};p_1,p_2)}{\partial p_1}&=\frac{({1\over e}p_1+(1-{1\over e})p_2)\log \frac{p_1}{{1\over e}p_1+(1-{1\over e})p_2}}{{1\over e}p_1+(1-{1\over e})p_2} -\frac{\frac{1}{e}(p_1-p_2)}{{1\over e}p_1+(1-{1\over e})p_2} \\
\frac{\partial^2 f(\frac{1}{e};p_1,p_2)}{\partial p_1^2}&=\frac{p_2\left((e-1)^2p_2-p_1\right)}{p_1\left(p_1+(e-1)p_2\right)^2}
\end{align}

\noindent Note that when $p_1=p_2$, $\frac{\partial f}{\partial p_1}=0$ and $\frac{\partial^2 f}{\partial p_1^2}\geq 0$. Therefore, for a fixed $p_2=c$, $p_1=c$ is a local minimum of the function $f(\frac{1}{e};p_1,p_2)$. Further, $f(\frac{1}{e};p_1,c)$ is convex for $p_1<(e-1)^2c$ and concave for $p_1>(e-1)^2c$.  Therefore, the global minimum of the function $f(\frac{1}{e};p_1,c)$ with respect to $p_1$ should occur either at $p_1=1$ or at $p_1=c$. Note also that $f(\frac{1}{e};c,c)=0$. Also, one can easily verify using calculus that $f(\frac{1}{e};1,c)\geq 0 ,~~\forall c\in[0,1]$. Thus, it follows that $f(\frac{1}{e};p_1,p_2)\geq 0$. \\ Lastly, it is easy to verify that the conditions of equality are $p_1=p_2$ and $p_2=0$.

\end{document}

%% file: defs.tex
\newcommand{\BEAS}{\begin{eqnarray*}}
\newcommand{\EEAS}{\end{eqnarray*}}
\newcommand{\BEQ}{\begin{equation}}
\newcommand{\EEQ}{\end{equation}}
\newcommand{\BIT}{\begin{itemize}}
\newcommand{\EIT}{\end{itemize}}

\newcounter{oursection}